\documentclass[12pt]{article}
\usepackage{amsmath}
\begin{document}
\begin{flushright}
DFF 261/12/96
\end{flushright}

\vspace{3cm}

\begin{center}

\Large{\textbf{Higher Topologies in 2+1-Gravity}}

\vspace{1.5cm}

\normalsize{Marcello Ciafaloni} 
\vspace{.5cm}

\small{\textit{Dipartimento di Fisica, Universit\`a di Firenze\\  
and I.N.F.N., Sezione di Firenze, Italy}}

\end{center}

\vspace{3cm}

\begin{abstract}
I argue that the first-order formalism recently found to describe
classical 2+1-Gravity with matter, is also able to include higher
topologies. The present gauge, which is conformal with vanishing
York time, is characterized by an analytic mapping from single-valued
coordinates to Minkowskian ones. In the torus case, this mapping
is based on four square-root branch points, whose location is
related to the modulus, which has a well defined time dependence.
In the general case, it is connected with the hyperelliptic
representation of Riemann surfaces.

\end{abstract}

\newpage

Classical solutions of (2+1)-Gravity \cite{car93} with dynamical
matter have recently been found \cite{bel95,wel96} in a (regular)
conformal gauge which allows instantaneous propagation of the
longitudinal gravitational field. This gauge is 
characterized by a vanishing York time \cite{han76}, and by a conformal
factor which is of Liouville type, even when arbitrary matter
sources are introduced.

Although the pointlike matter described so far is characterized
by a non-compact conical geometry at space infinity, corresponding
to a total mass of an open universe, in this note I suggest that
the method can be extended to some compact cases.
Furthermore, I will argue, it may describe proper matter
sources as well, possibly combining a non-compact structure with
higher topologies. 

To set up the present method, I will describe
in detail the torus case, that was solved long ago by Moncrief 
\cite{mon89}
in the (non vanishing) York time gauge. (2+1)-Gravity on a torus
is characterized by two nontrivial Poincar\'e holonomies around
the two homotopy cycles, described by the DJH \cite{des84}
matching conditions
\begin{equation}
X^{a})^{II} - B^{a}_{\alpha} = \Lambda^{a}_{b} (\lambda_{\alpha})
(X^{b})^{I} - B^{b}_{\alpha}) \; , (\alpha=1,2) \; .
\label{XaII}
\end{equation}
where $X^{a})^{I} (X^{a})^{II})$ denote the (multivalued)
Minkowskian coordinates before (after) application of the
holonomies. Such Poincar\'e transformations are constrained to 
commute, because
the compound holonomy around the period parallelogram is trivially
contractible, and their Lorentz parts can be chosen to be $x$-boosts
$\lambda_{1},\lambda_{2}$.

The mapping from single-valued coordinates $x^{\mu} \equiv (t,z,
\bar{z})$ to Minkowskian ones $X^{a} \equiv (T,Z,\bar{Z})$ is best
expressed in the first-order form
\begin{equation}
dX^{a} = E^{a}_{\mu} (t,z,\bar{z}) dx^{\mu}
\label{dXa}
\end{equation}
where the dreibein $E^{a}_{\mu}$, because of 
(1),
is multivalued and satisfies the monodromy conditions
\begin{equation}
E^{a}_{\mu})^{II} = \Lambda^{a}_{b} (\lambda_{\alpha})
E^{b}_{\mu})^{I} \;\; , (\alpha=1,2) \; ,
\label{Eamu}
\end{equation}
along the homotopy cycles, while the metric $g_{\mu\nu} =
E^{a}_{\mu}E_{a\nu}$ is not affected by Eq~(\ref{Eamu})
and is thus single-valued.

It is convenient to use an ellyptic representation \cite{coh67} 
of the torus,
and to introduce the $u$ coordinate, characterized by the
holomorphic differential
\begin{equation}
du = \frac{dz}{w(z)} \; , \; w^{2} = 4 z (z-1)(z-a(t)) \; ,
\label{du}
\end{equation}
which provides the customary representation of the torus
on the two-sheeted $z$-plane, in terms of 4 branch-points
at $z=0,1,a(\tau),\infty$,where
\begin{equation}
\tau = \frac{\omega_{1}}{\omega_{2}} = \frac{1}{\sqrt{a}} \;
\frac{F(\frac{1}{2},\frac{1}{2};1;\frac{1}{a})}
{F(\frac{1}{2},\frac{1}{2};1;a)}
\label{tau}
\end{equation}
is the modulus and $2\omega_{1}, 2\omega_{2}$ are the torus'
periods.

By injecting in Eq~(\ref{dXa}) the no-torsion and conformal conditions 
$(g_{zz}=g_{\bar{z}\bar{z}}=0)$ and the Coulomb condition of
vanishing York time $K \sim \partial_{z} E^{a}_{\bar{z}} +
\partial_{\bar{z}} E^{a}_{z} =0$, we find that $E^{a}_{z}
(E^{a}_{\bar{z}})$ is analytic (antianalytic) while $E^{a}_{0}$
is harmonic, with the null-vector representations
\begin{equation}
E^{a}_{z} = \frac{N(z,t)}{f'(z,t)} \left(
\begin{array}{c}
f \\ 1\\ f^{2}
\end{array} \right) \; , \;\;
E^{a}_{u} = \frac{n(u,t)}{\dot{f}(u,t)} \left(
\begin{array}{c}
f \\ 1\\ f^{2}
\end{array} \right)
\label{Eaz}
\end{equation}
and the notation
\begin{equation}
N(z(u),t) = \frac{n(u)}{w^{2}(z)} \; , \;\;
\dot{f} = \frac{df}{du} = w(z) f'(z,t) \; .
\label{Nzu}
\end{equation}

Then the vector monodromies of $E^{a}_{u}$ in Eq~(\ref{Eamu})
are satisfied by imposing projective monodromies on the
mapping function $f$ in Eq~(\ref{Eaz}). Since the $\lambda$'s
are $x$-boosts, we can satisfy Eq~(\ref{Eamu}) by setting
\begin{eqnarray}
f &=& th \Theta (u) \; , \nonumber \\
2\Theta (u+2\omega_{\varrho}) &=& 2\Theta (u) +  
\lambda_{\varrho} \; , ( {\rm mod}\, 2\pi {\it ni}, \varrho =1,2) \; .
\label{fth}
\end{eqnarray}

Then, $n(u)$ by Eq~(\ref{Eaz}) and $\dot{\Theta}(u)$ by
Eq~(\ref{fth}) are meromorphic functions on the torus, i.e.
ellyptic or biperiodic functions of the $u$ variable.
Notice that $N(z)$ turns out [2] to be one of the components of the 
extrinsic curvature, the other one being the (vanishing) York time.

A solution to Eq~(\ref{fth}) entails, by  Eq~(\ref{Eaz}),
a solution to Eqs~(\ref{dXa}) and (\ref{Eamu}), which for the
combinations $X_{\pm} = T \pm X, X_{2} = Y$ take the form
$(dt=0)$
\begin{equation}
dX_{\pm} = \pm du \frac{n}{2\dot{\Theta}} e^{\pm 2\Theta} + c.c. \; ,
\; dY = du \frac{n}{2i\dot{\Theta}} + c.c. \; ,
\label{dXpm}
\end{equation}
where we can assume $n$ and $\dot{\Theta}$ to be even in $u$, so that
$Y$ and $\Theta$ are odd, and $X_{-}(u) = X_{+}(-u)$. By solving
for $\Theta$ and $n$, we can integrate Eq~(\ref{dXpm}) and impose
the full DJH matching conditions in Eq~(\ref{XaII}), including the
translational part.

The ellyptic function $\dot{\Theta}(u)$ is in general given \cite{coh67}
by the ratio of two polynomials in $z \equiv P(u)$, the even
Weierstrass ellyptic function. The simplest nondegenerate solution
is an ellyptic function of order 2, namely
\begin{equation}
2\dot{\Theta}= A + \zeta(u-\alpha) - \zeta(u+\alpha) \sim \tilde{A}
\; \frac{z(u)-\eta}{z(u)-\nu} \; ,
\label{2Theta}
\end{equation}
where we have introduced the notation
\begin{eqnarray}
z(u) &=& P(u) = -\dot{\zeta}(u) \; , \nonumber \\
\nu &=& z(\alpha) = z(-\alpha)
\label{zu}
\end{eqnarray}
and the $\zeta$-function satisfies the monodromies
\begin{eqnarray}
\zeta(u+2\omega_{\varrho}) &=& \zeta(u) + 2\gamma_{\varrho} \; , \nonumber \\
\gamma_{1}\omega_{2}- \gamma_{2}\omega_{1} &=& \frac{1}{2} \pi i \; .
\label{xiu}
\end{eqnarray}

Eq~(\ref{2Theta}) shows simple pole singularities at $u=\pm
\alpha (z=\nu)$, which however are harmless, because the 
integrated $\Theta$ variable
\begin{equation}
2\Theta(u) = Au + \log \frac{\sigma (u-\alpha)}
{\sigma(u+\alpha)} \; , \; \;
\left(\zeta=\frac{\dot{\sigma}(u)}{\sigma(u)} \right) \; ,
\label{2Thetau}
\end{equation}
changes by $\pm 2\pi i$ when turning around $u=\pm \alpha$,
thus leaving $\exp (\pm 2\Theta)$ invariant.

Then, by imposing the monodromies in Eq~(\ref{fth}) we get
the conditions
\begin{equation}
2\omega_{\varrho} A-4\gamma_{\varrho}\alpha = \lambda_{\varrho}
\;, \;\; (\varrho = 1,2) \; ,
\label{2omega}
\end{equation}
which, by exploiting Eq~(\ref{xiu}), determine $A$ and $\alpha$
as functions of $\lambda_{1}$ and $\lambda_{2}$:
\begin{equation}
\pi i A = \gamma_{1}\lambda_{2} - \gamma_{2}\lambda_{1} \; ,
\;\; 2\pi i \alpha = \omega_{1}\lambda_{2}-\omega_{2}\lambda_{1}
\; .
\label{piiA}
\end{equation}

Note that $\alpha$ vanishes in the degenerate limit
$\tau = \omega_{1}/\omega_{2} \rightarrow \lambda_{1}/
\lambda_{2}$ (real), in which case the pole at $z=\nu$
in Eq~(\ref{2Theta}) is avoided, and $\dot{\Theta}$
reduces to a constant.

In the general case, $\Theta$
in Eq~(\ref{2Thetau}) has a logarithmic singularity at
$u = \pm \alpha$, while $e^{\pm 2\Theta}$ shows a pole
in the form
\begin{equation}
e^{\pm 2\Theta} = e^{\pm Au} \; \frac{\sigma(u\mp \alpha)}
{\sigma(u \pm\alpha)} \; .
\label{epm2}
\end{equation}
In order to cancel this pole in Eq~(\ref{dXpm}) we shall set
\begin{equation}
n=\frac{1}{2} C(t) \tilde{A} (z(u)-\eta) \; , \;\;
\frac{n}{\dot{\Theta}} = C(t) (z(u)-\nu)
\label{n12}
\end{equation}
where $\eta$ is a common zero of $n$ and $\dot{\Theta}$ and is
thus \cite{bel95} an ``apparent singularity''\cite{yos87}, 
and the normalization
constant $C$ will be determined shortly.

Having determined the form of $n$ and $\dot{\Theta}$ in Eqs~(\ref{2Theta}) and
(\ref{n12}), we can calculate the Schwarzian derivative
\begin{equation}
\{ f,z\} = \frac{1}{\omega^{2}} \left( \{\Theta,u\} 
-2\dot{\Theta}^{2}\right) + \{u,z\} \; ,
\label{fz}
\end{equation}
which provides the potential of the related Fuchsian problem
\cite{yos87}. The Schwarzian turns out to have 5 singularities,
the normal ones at $z=0,1,a(\tau),\infty$ with common difference
of exponents $\mu_{\alpha} =\frac{m_{\alpha}}{2\pi}=\frac{1}{2}$, and
the apparent singularity at $z=\eta$, which has $\mu_{5}=2$ and trivial
monodromy. The one at $z=\nu$ is instead absent altogether, 
and appears to be needed 
only in the intermediate steps of the construction.

The integration of Eq~(\ref{dXpm}) now proceeds without
troubles. Note first that the only singularity of the $X$'s
comes from the double pole at $u=0$, or $z=\infty$, appearing
in the expressions
\begin{eqnarray}
\frac{dX_+(u)}{du} &=& Ce^{Au} \; \frac{\sigma^{2}(u-\alpha)}
{\sigma^{2}(\alpha) \sigma^{2}(u)} \; + c.c. \; , \nonumber \\
\frac{dY(u)}{du} &=& \frac{1}{i} C (z(u)-\nu) + c.c. \; ,
\;\; dX_{-}(u) = dX_{+}(-u) \; .
\label{dXu}
\end{eqnarray}
Expanding around $u=0$ we get the behaviour
\begin{equation}
X(u) + i Y(u) \simeq - \frac{C}{u} \sim \sqrt{z(u)} \; ,
\;\; (u\rightarrow 0) \; ,
\label{Xui}
\end{equation}
which provides the same square-root behaviour as for the
other branch-points. This checks with the previous result that the 
``particle masses'' at $z=0,1,a,\infty$ are all equal to 
$\pi$.

This remark shows that the handle of the torus is here obtained 
from quantized
pointlike singularities of the extrinsic curvature $N(z)$ in
Eq~(\ref{Nzu}). 
Furthermore,
the present mapping wraps the torus on an infinite 2-dimensional 
slice of $X$-space,
rather than on a bounded ``cell''.

As a second remark, the equations of motion for $C(t)$ and
$\tau(t)$ follow from the translational part of the DJH matching
conditions in Eq~(\ref{XaII}). While in the particles' 
case it is natural to measure time by the clock of one of them,
in the present case we just choose to set, $X_{+}(-\omega_{1}
-\omega_{2})=t$, in a somewhat arbitrary way.

Solving for $Y$ in Eq~(\ref{dXu}) is simple, and yields
\begin{equation}
Y(u)=-ImC(\zeta(u)+\nu u)
\label{Yu}
\end{equation}
so that the translational monodromy in Eq~(\ref{XaII}) reads
\begin{equation}
Im2C(\gamma_{\varrho} +\nu\omega_{\varrho})=-B_{\varrho}
\; , \;\; (\varrho=1,2) \; .
\label{Im2C}
\end{equation}

Integrating for $X_\pm$ is less explicit and yields
\begin{equation}
X_{+}(u) = t -Re\left(C \int^{u}_{-\omega_{1}-\omega_{2}}
du \, e^{Au} \; \frac{\sigma^{2}(u-\alpha)}
{\sigma^{2}(\alpha)\sigma^{2}(u)}
\right)\; ,
\label{X+u}
\end{equation}
so that Eq~(\ref{XaII}) provides, after simple algebra, the
equations
\begin{eqnarray}
&& \frac{1}{1-e^{\lambda_{1}}} Re \left(
C \int^{\omega_{1}-\omega_{2}}_{-\omega_{1}-\omega_{2}}
du\, e^{Au} \; \frac{\sigma^{2}(u-\alpha)}
{\sigma^{2}(\alpha)\sigma^{2}(u)}
\right) = k-t = \nonumber \\
&=& \frac{e^{\lambda_{1}}}{1-e^{\lambda_{2}}} 
Re \left(
C \int^{\omega_{1}+\omega_{2}}_{\omega_{1}-\omega_{2}}
du\, e^{Au} \; \frac{\sigma^{2}(u-\alpha)}
{\sigma^{2}(\alpha)\sigma^{2}(u)}
\right)\; .
\label{11e}
\end{eqnarray}

Eqs~(\ref{Im2C}) and (\ref{11e}) provide four real conditions
which determine the complex parameters $C$ and $\tau$ as functions
of time. For instance, in the ``quasistatic'' limit
$|\lambda_{\varrho}t| \ll B_{\varrho} , \alpha =0(\lambda_{\varrho})$ in
Eq~(\ref{piiA}) is small, and Eqs~(\ref{Im2C}) and (\ref{11e}) simplify
to
\begin{equation}
Im(2C\nu\omega_{\varrho}) = -B_{\varrho} \; , \;\;
Re(2C\nu\omega_{\varrho} = \lambda_{\varrho}(t-k) \; ,
\label{Im2}
\end{equation}
thus yielding the expression
\begin{equation}
\tau(t) = \frac{B_{1}+i\lambda_{1}(t-k)}{B_{2}+i\lambda_{2}(t-k)}\; .
\label{taut}
\end{equation}
which describes a circle, as in Moncrief's solution \cite{mon89}.

For finite $\lambda$ values, however, changing from the York time gauge
to the Coulomb gauge mixes analytic with antianalytic functions, so that
the modulus trajectory is here more complicated than a circle.

Let me now come to the issue of introducing pointlike matter. This can
be done either by keeping the compact topology, or by introducing a
boundary as well. In the first case, since the holonomy around the
period parallelogram is still trivially contractible, we need at least
two particles - for instance, static ones with masses $m$ and $4\pi-m$,
- so as to yield a trivial compound holonomy. A branch-cut will join the
particles and the solution for $f$ or $\Theta$ will be found by solving
a Fuchsian problem with nontrivial boundary conditions at the edges of
the period parallelogram.

The general case of pointlike matter requires a boundary, which without
loss of generality can be set around $u=0 (z=\infty)$ by having all cuts
terminating at that point. In this case the trivial contractibility
argument relates the holonomy around the boundary to both particle'
momenta and torus' boosts, much in the same way as it happens in the
multiconical geometry \cite{bel95}.

Finally, higher genus topology could be treated following the same
method as for the torus. The basic idea is that we could add a handle by
adding particles with quantized momenta, - usually not in a unique way -
, so as to fit the compact surface requirements. We should therefore get
some representation of the higher genus gravitational problem which is
much similar to the hyperelliptic ones of Riemann surfaces, with the
possible addition of some apparent singularities.

The actual construction of such solutions is still under investigation.
\vspace*{.5 cm}

\textbf{Acknowledgements}
\vspace*{.5 cm}

I wish to thank Alessandro Bellini, Steve Carlip, 
Renate Loll, Pietro Menotti and
Paolo Valtancoli for interesting discussions and comments. 
This work is supported in part by M.U.R.S.T. (Italy).

\end{document}